\documentclass[11pt, a4paper]{article}
\pdfoutput=1
\usepackage{jheppub}
\usepackage{graphicx}
\usepackage{epsfig,amsmath,amsthm}
\usepackage{epstopdf}

\newcommand{\be}{\begin{equation}}
\newcommand{\ee}{\end{equation}}
\newcommand{\bea}{\begin{eqnarray}}
\newcommand{\eea}{\end{eqnarray}}
\def\bse{\begin{subequations}}
\def\ese{\end{subequations}}
\newcommand{\IR}{\mathbb{R}} 
 
\def\IZ{\relax\ifmmode\hbox{Z\kern-.4em Z}\else{Z\kern-.4em Z}\fi}

\newcommand{\non}{\nonumber \\}

\def\half{\frac{1}{2}} 

\def\del{{\partial}}

\def\presub{\vspace{.5cm} \noindent}

\def\bi{\begin{itemize}} \def\ei{\end{itemize}}

\def\({\left(} \def\){\right)}
\def\[{\left[} \def\]{\right]}
\def\<{\left<} \def\>{\right>}

\def\tI{\widetilde{I}}

\title{Symmetries of Feynman integrals\\ and the Integration By Parts method}

\author{Barak Kol  \\
{\it Racah Institute of Physics, Hebrew University, Jerusalem 91904, Israel} \\
{\tt barak.kol@mail.huji.ac.il}
}

\abstract{Integration By Parts (IBP) is an important method for computing Feynman integrals. This work describes a formulation of the theory involving a set of differential equations in parameter space, and especially the definition and study of an associated Lie group $G$. The group acts on parameter space and foliates it into $G$-orbits. The differential equations essentially provide the gradient of the integral within the orbit in terms of integrals associated with degenerate diagrams. In this way the computation of a Feynman integral at a general point in parameter space is reduced to the evaluation of the Feynman integral at some freely chosen base point on the same orbit, together with a line integral inside the $G$-orbit and the degenerate integrals along the path. This paper restricts to vacuum diagrams and integrals without numerators, but the method is expected to generalize. The method is demonstrated by application to the two-loop vacuum diagram. A relation between the reducibility of a diagram though IBP and the reducibility of the associated electric circuit though the $Y-\Delta$ transform is speculated.} 
 
\begin{document}
\maketitle

\begin{flushright}
To the reader
\end{flushright}


\section{Introduction}
\label{sec:intro}

\begin{figure}[b!]
\centering \noindent
\includegraphics[width=5cm]{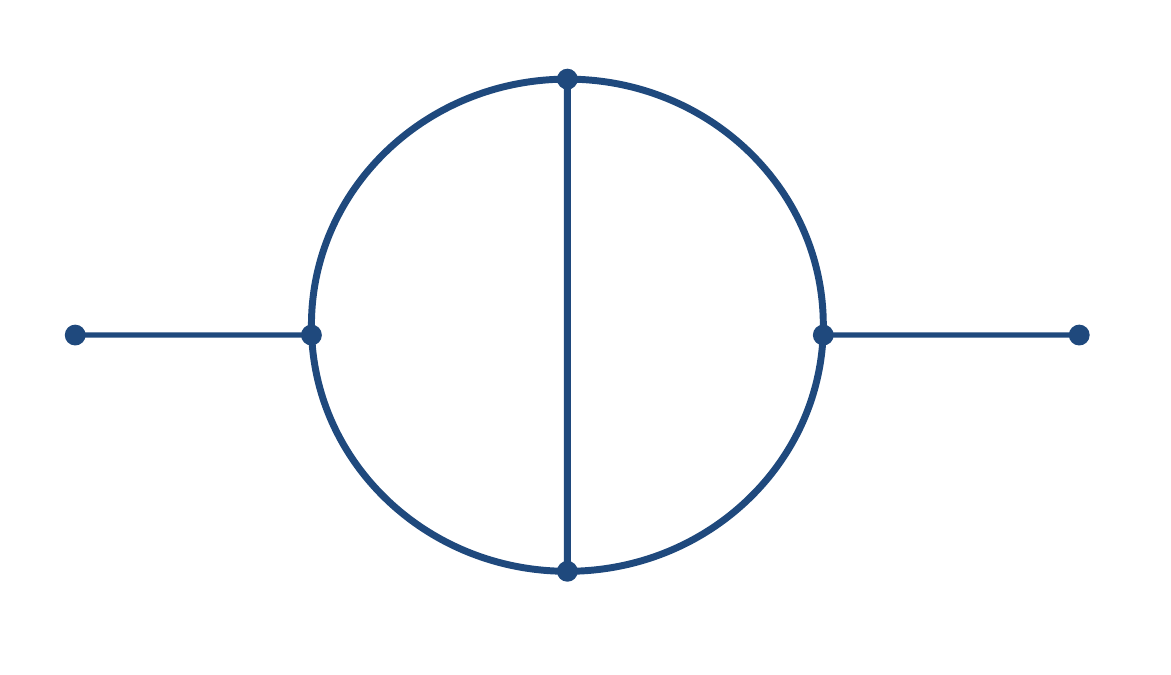}
\caption[]{The two-loop propagator diagram. The associated Feynman integral is conveniently computed by the IBP method.}
 \label{fig:Ln22}
\end{figure}

Integration By Parts (IBP) is an important method for computing Feynman integrals, discovered in 1981 \cite{ChetyrkinTkachov81}. It should be distinguished from the general method of effecting integration which bears the same name. IBP enables to compute many diagrams, for instance the 2-loop propagator diagram shown in fig.\ref{fig:Ln22}, and whenever it applies it works  for all space-time dimensions. Ref.  \cite{ChetyrkinTkachov81} currently has about one thousand InSpire citations, evidence for both its wide application and for its theoretical interest.

Despite its popularity and 34 years of research, so far it is not known to which diagrams the method should be applied and what output should be expected of it. Already the original authors remarked  (in the discussion section) ``As yet there is no criterion to decide whether a given 4 loop $p$-integral [propagator type integral - BK] can reduced to simpler ones via integration by part''  as well as ``All these observations leave the impression of something important having been missed in our analysis, that could be very useful for both practical purposes and a better understanding of perturbative series, if these two things can be separated'' . Indeed this question is natural for any method, a sort of `declaration of purpose' to appear in the `instruction manual'. (It is hoped that the quotations would not trigger the arXiv `text overlap' alert...)

We can mention here a few more specific indicators for the incompleteness of the current theory for the IBP method. Presently the practice is to use Laporta's  algorithm \cite{Laporta2001} realized through computer packages both for  obtaining the IBP recursion relations, and then for solving the recursion. Indeed, computers are very useful to carry long computations, yet blind usage of `black boxes' may obscure the general properties of the method.  In addition, the very name of the method, I find to be lacking, since it shares the name of the general method of integration, while the potential of IBP depends strongly on the diagram topology, and that is not reflected by the name. This is another indication for an incomplete state of understanding. 

In this paper I shall carry some steps to discover the `statement of purpose', and I believe the full answer will be found soon. Section \ref{sec:generating} defines a generating function for Feynman integrals for a given diagram topology,  and its computation would be our goal.  Section \ref{sec:IBP} describes how to obtain IBP relations both in index space $n$ and in mass-squared space $x$. The main novelty appears in section \ref{sec:Lie} where we motivate and define the numerator-free sub-algebra of IBP relations, which leads to a system of linear first order partial differential equations. 
We proceed to outline the general procedure of solving the equations in section \ref{sec:soln}. Next in section \ref{sec:demonstrate} we demonstrate the ideas of this paper through a concrete diagram -- the two loop vacuum diagram. Finally we offer a summary and discussion in section \ref{sec:summary}.


Being a newcomer to the subject the author may have omitted relevant references, and he would be happy to receive correspondence on this and other matters.


\presub {\bf Background}. The IBP method was discovered in \cite{ChetyrkinTkachov81}. Baikov's method, a (non-recursive) solution to the recursion relations, was presented in 1996 \cite{Baikov1996}.  During the 90's the closely related method of Differential Equations (DE) was  developed \cite{Kotikov1990,Remiddi1997,GehrmannRemiddi1999}, see the reviews \cite{ArgeriMastrolia2007,HennRev2014} where the last one also reviews recent progress. The IBP and DE methods occupy a prominent place in the excellent textbooks \cite{SmirnovBook2006,SmirnovBook2012} on Feynman integrals. Fairly recent works on IBP include: an introduction to IBP \cite{Grozin2011}; a comparison of master integral counting between IBP and DE \cite{Kalmykov:2011yy}; an algorithm for the generation of unitarity-compatible IBP Relations \cite{Schabinger:2011dz}; new relations between master integrals using IBP \cite{Kniehl:2012hn};  a recent version of IBP computerized tools \cite{Smirnov2013dia}; an improvement over Laporta's algorithm which employs a Monte-Carlo approach \cite{Kant:2013vta};  a suggested improvement for Laporta's algorithm using finite fields \cite{vonManteuffel:2014ixa}; a recent work from the viewpoint of differential geometry utilizing differential forms \cite{Zhang2014}; and finally an application to multiploop Four Dimensional Regularized integrals \cite{Pittau:2014tva}.

The author learned about the IBP method through its application to the analysis of the post-Newtonian limit of the two-body problem in Einstein's gravity  \cite{GilmoreRoss2008,KolShir2013} (which is a non-quantum problem).

\section{Generating function}
\label{sec:generating}

Given a Feynman diagram $\Gamma$ we would like to compute a general integral of the form \be
I_\Gamma(x,n) := \int \frac{dl}{\prod_i \( k_i^2-x_i \)^{n_i}}  
 \label{def:Ixn}
\ee
where $dl := \prod_{r=1}^L d^dl_r$ denotes integration over all loops for a general space-time dimension $d$; the index $i=1,\dots,P$ runs over all propagators; $k_i^\mu$ is the energy-momentum of the propagator and it should be considered to be expressed as a combination of loop and external momenta; the $x$ parameters are standard notation and can be identified with squared masses $x_i \equiv m_i^2$; and finally the powers $n_i$ are termed indices. $I_\Gamma(x,n)$ is defined as a very general Feynman integral associated with $\Gamma$. The indices $n_i$ generalize the common case $n_i=1$ and are necessary as they may appear during IBP manipulations of unit indices.  More generally the integral could also include numerators with arbitrary powers of Lorentz scalars, but for the current purposes (\ref{def:Ixn}) suffices.


The general integrals $I_\Gamma(x,n)$ for all natural indices $n_i=1,2,3,\dots$ can be conveniently encoded into the generating function \be
I_\Gamma(x) := \int \frac{dl}{\prod_i \(k_i^2-x_i\)}  \equiv I_\Gamma(x,n) |_{n_i=1}
 \label{def:Ix}
\ee
Indeed $I(x,n)$ can be recovered through  \be
I(x,n)=\( \prod_i \frac{1}{(n_i-1)!)} \(\frac{\del}{\del x_i}\)^{n_i-1} \) \, I(x) 
\ee
Since $I(x,n)$ is encoded in the Taylor coefficient of $I(x)$ it is appropriate to refer to the latter as a generating function. Note that whereas an ordinary generating function encodes a single series of coefficients through its Taylor coefficients at one specific point, such as the origin, here the Taylor coefficients of $I(x)$ around \emph{any} point are of interest. In addition, since $I(x) = I(x,n_i=1)$ this generating function has a very clear interpretation, and the $x$ variables carry a double interpretation -- both as formal parameters of the generating function as well as mass squares. Hence we find that the integrals $I(x)$ contain as much information as $I(x,n)$ and we set a goal to compute  $I_\Gamma(x)$.


\section{IBP relations and generators}
\label{sec:IBP}


The standard way to derive the IBP relations is to consider the identity \be
 0 = \int dl \, \frac{\del}{\del l} \, k \, \tI 
\label{IBP-gen}
\ee
where $\del/\del l \equiv \del/\del l^\mu$ is a divergence with respect to one of the loop momenta; $k \equiv k^\mu$ is a propagator momentum and $\tI$ is any integrand, which we will mostly take to be the integrand of (\ref{def:Ixn}), namely \be
\tI =  \frac{1}{\prod_i \( k_i^2-x_i \)^{n_i}} ~.
\label{def:tI}
\ee 
The vector indices $\mu$ are suppressed for clarity, and more generally $k$ could be a linear combination of propagator momenta, or equivalently of loop and external momenta. 

By Gauss's divergence theorem the RHS of (\ref{IBP-gen}) equals a flux integral over a surface at infinity. For small enough dimensions this integral vanishes (in the limit that the surface approaches infinity) and hence in dimensional regularization, which is assumed here, it vanishes identically. The method's name is derived from the use of the divergence theorem which is a generalization of the elementary integral identity of integration by parts.\cite{ChetyrkinTkachov81} adds another point of view, that of configuration space, explaining that the divergence identity (\ref{IBP-gen}) is related to the vanishing sum of 3 vectors representing triangle edges (eq. 3.4b over there).

In order to obtain the IBP relations we proceed to expand the RHS of (\ref{IBP-gen}). When $k$ lies in the loop $l$ its differentiation generates a term with a factor of $d$. Next we should differentiate the propagators in the denominator. For any $k_i$ lying in the loop $l$ we get a term of the form \be
 2 k \cdot k_i \, n_i\, {\bf i}^+ 
 \ee
 where ${\bf i}^+$ is a raising operator which increases the integrand's index $n_i$ by 1.
 
Now comes a crucial point. The Lorentz scalar $k \cdot k_i$ can be expanded as \be
2 k \cdot k_i = \sum_j T^i_j\, k_j^2 + M_i ~,
\ee
namely into a combination of squares $k_j^2$ plus possibly a remainder, denoted $M_i$. Such factors can appear in the numerators of more general Feynman integrals. For example, in a trivalent junction where $k_2=k-k_1$ we have \be
2 k \cdot k_1 = k^2 + k_1^2 - k_2^2 
\ee

Defining a propagator term \be
E:= k^2 - x
\label{def:x}
\ee
we can replace $k^2_j = E_j + x_j$ and then we further replace $E_j \to {\bf j}^-$, where ${\bf j}^-$ is a lowering operator which reduces $n_j$ by 1. Finally we arrive at the following types of terms which can appear upon expanding the RHS of (\ref{IBP-gen}) \be
d    ~,~    n_i\, {\bf i}^+ \( {\bf j}^- + x_j \)    ~,~    n_i\, {\bf i}^+\, M_i
\label{IBP-terms}
\ee

The operators ${\bf i}^+,\, {\bf i}^-$ relate integrals with different values of $n_i$ and hence we obtained a recursion relation for $I$. This relation is clearly linear in $I$, and it is of order at most 1 both in raising and in lowering operators. 
In fact, by choosing different values for $l$ and $k$ we obtain a set of linear recursion relations, known as the IBP relations. More specifically for a vacuum diagram with $L$ loops we may generating an independent set of equations by choosing $l=l_r,\, k=l_s$ where $r,s=1,\dots,L$ run over all loops.

\subsection{IBP relations in $x$ space}

We notice that the terms appearing in the IBP recursion relations (\ref{IBP-terms}) are special. A term of the form $f(n_i)\, {\bf i}^+ {\bf j}^-$ with $f(n_i)$ an arbitrary function of $n_i$ would preserve linearity and the order of the IBP recursion relation. However, we notice that the dependence on $n$ appears in a special form $n_i\, {\bf i}^+$, namely, a factor of $n_i$ is always accompanied by a raising operator. 

This leads us to seek a set of variables where this property may have a more natural interpretation.  In general it is known that recursion relations are connected with differential equations: a differential equation becomes a recursion equation once we substitute in its Taylor series expansion. Conversely, a recursion relation can be equivalent to a differential equation for its generating function, in some formal parameters $x$. In this way we could motivate the definition of the generating function (\ref{def:Ix}) even if we started with completely massless integrals in $n$ space, namely $I(x_i=0,n_i)$. In fact one can consider various ways of collecting a series $I(n)$ into a generating function $I(x):=\sum_{n=1}^{\infty} f(n)\, I(n)\, x^{n-1}$ where $f(n)$ could be an arbitrary function of $n$, such as $f=1$ or $f=1/(n-1)!$. The choice (\ref{def:Ix}) corresponds to choosing $f(n)=1$ and is exactly such that the factor $n_i\, {\bf i}^+$ is replaced by a derivative $\del/\del x_i$.

An IBP relation in $x$ space is derived by starting with the same identity (\ref{IBP-gen}) where this time all indices are set to unity in the integrand $\tI$ (\ref{def:tI}). Whereas in the recursion relation context the mass squares $x_i$ were considered fixed, here we notice that we can replace $n_i\, {\bf i}^+ \to \del/\del x_i$ as anticipated. The appearance of $\del/\del x_i$ can be thought to arise from the identity \be
 \frac{\del}{\del k_i^2} =-\frac{\del}{\del x_i}
\ee
 which holds for all functions of the form $\tI=\tI(k_i^2-x_i)$. Now the derivative $\del/\del x_i$ can be taken outside the integral, and then an IBP relation can be stated at the level of the integral $I(x)$ as follows \be
0 = \[ c + T^i_{~j}\, x_i \, \del^j \] I + J
\label{IBPx}
\ee
where $c$ and $T^i_{~j}$ are constants ($c$ may depend on $d$ as follows $c(d)=c_1\, d - c_2$) , $\del^j \equiv \del/\del x_j$ and finally $J$ is a source term  \be 
J = \int dl \[ T^i_{~j}\, \del^j \, E_i + M_j  \, \del^j \] \tI \nonumber
 \ee
 where $M_j$ are  some numerators. Noticing that \be
  \int dl \, E_i \,  \tI = O_i I \ee
  where $O_i I_\Gamma:= I_{O_i \Gamma}$ and $O_i \Gamma $ is the diagram obtained from $\Gamma$ by omitting propagator $i$, the first term of $J$ can be written in terms of generating functions for degenerated diagrams, namely \be
  J = T^i_{~j}\, \del^j \, O_i  I+  \del^j \int dl \, M_j  \,  \tI ~ . 
  \label{def:J} \ee
 
The differential equations appear to be the same at those suggested by \cite{Tarasov1998}, and are also related to the approach of \cite{Baikov1996}.  
  
Altogether in $x$ space we arrive at the differential equation (\ref{IBPx}) with the source term (\ref{def:J}). If there are several IBP relations labelled by $a$ then we obtain a set of differential equations and their parameters obtain an $a$  index as follows \be
0 = \[ c^a + \(T^a\)^i_{~j}\, x_i \, \del^j \] I + J^a
\label{IBPxa}
\ee
 where $J^a$ depends now on $M^a_j$. 
  
 These equations are linear in $I$ and first order in the derivatives $\del/\del x_i$. It should be stressed that the first degree in derivatives is not a direct consequence of the first degree of the recursion relations, but rather it relies on the special dependence on $n_i$ which we noticed. In this sense the IBP relations are more natural in the $x$ plane. 
  
\section{Lie algebra structure and sub-algebras}
\label{sec:Lie}


The IBP equations in $x$-space (\ref{IBPxa}) immediately suggest a Lie algebra structure (it is enough that the equations are linear and of first order in derivatives). It is convenient to define differential operators \be
D^a := c^a  + \(T^a\)^i_{~j}\, x_i \, \del^j 
\label{def:D}
\ee
such that the differential equations become (\ref{IBPxa}) \be
 0=D^a\, I + J^a
 \label{IBPxD}
 \ee 
Consider the commutator $[D^a,D^b]$. The terms quadratic in derivatives cancel, so the commutator is linear in derivatives. As such it generates another equation of the same type. If we already had all the equations, then this commutator must be linear combination of the differential operators, namely \be
\[  D^a,D^b \] = f^{ab}_{~~c} D^c ~.
 \label{comm-relationsD}
\ee
This is a Lie algebra structure with  structure constants $f^{ab}_{~~c}$. 

We can be a bit more explicit about the commutation relations of the differential operators (\ref{def:D}) \be
\[ D^a, D^b \] = \( \[ T^a ,T^b \] \)^i_{~j} x_i\, \del^j
\ee
where the commutator on the LHS is the operator commutator and on the RHS it is simply the matrix commutator.

The Lie algebra structure of the IBP relations was described already by R. Lee in  \cite{Lee2008}. There the algebraic structure was studied in relation to the Lorentz Invariance identities for external momenta. It was described in $n$ space and studied through the commutators of the IBP generators $O_{rs}= \del/\del l_r\, l_s$, which are the same operators appearing in (\ref{IBP-gen}). Clearly the group structure among the differential operators $D^a$ (\ref{comm-relationsD}) is inherited from  the commutation relations of the associated $O^a$ operators (namely, the mapping $O \to D$ preserves the Lie algebra structure).
 That paper does not mention the $x$ space form of the IBP equations, where as we have shown the Lie group structure is apparent.

Applying the commutator to the IBP equations (\ref{IBPxD}) we obtain a constraint. We start with \bea
 0 &=& D^a \( D^b\, I + J^b \) - D^b \( D^a\, I + J^a \) = \non
  &=& \[ D^a, D^b \]\, I + D^a\, J^b - D^a\, J^b
\eea
Using (\ref{comm-relationsD}) in the last line and comparing with $0=f^{ab}_{~~c} \( D^c\, I + J^c \)$ we now obtain \be
 D^a\, J^b - D^a\, J^b = f^{ab}_{~~c} J^c
 \label{integ-constraint}
\ee
This is an integrability constraint which the sources $J^a$ must satisfy.

\subsection{The numerator-free sub-algebra}

{\bf Defintion}. We denote by $G_0$ the Lie algebra of all IBP relations, namely all the operators $D^a$ (\ref{def:D}) (and the associated equations (\ref{IBPxa}) ) obtained from all possible identities of the form (\ref{IBP-gen}) for arbitrary $l$ and $k$ vectors (in the presence of external legs these identities should be supplemented \cite{Baikov1996,BaikovSmirnov2000}). 

Considering the source term (\ref{def:J}) we observe that the first term contains degenerations of $\Gamma$ and hence will be assumed to be known (otherwise we have no right to expect to be able to compute $I_\Gamma$) while the second term includes a more complicated and therefore possibly unknown integral. This motivates the following definition:

\presub {\bf Definition}. We denote by $G$ the subspace of $G_0$ such that all sources (\ref{def:J}) are free of numerators. 

A similar motivation can be seen in $n$ space. In the presence of numerators the IBP relations contains terms which increase an index in the numerator (as well as the index of a denominator), and hence the resulting recursion relation could be considered ineffective, as the integral with the numerator would be usually unknown. 

$G$ is not only a sub-space but rather a sub-algebra of $G_0$, as the commutator of two numerator-free relations is again numerator-free. We refer to $G$ as the \emph{numerator-free Lie algebra}. 

From now on we restrict ourselves to vacuum diagrams and integrals without numerators (\ref{def:Ix}). Including numerators is possible, but as we have seen it is necessary first to solve for the numerator-free integral. Including diagrams with external legs complicates the discussion a bit, and so incorporating it in the current formulation is reserved for future work, though in principle it is understood \cite{Baikov1996,BaikovSmirnov2000}).

Let us summarize the set of equations for $I(x)$ associated with $G$ \be
0 = D^a\, I + J^a  \label{def:set}
\ee
where \bea
D^a &=& c^a + \(T^a\)^i_{~j}\, x_i \, \del^j \non
J^a &=&  \(T^a\)^i_{~j}\, \del^j \, O_i  I  \nonumber
\eea
where $a$ runs over the generators of $G$. This is a set of linear first order PDE's. Moreover, the operators $D^a$ (\ref{def:D}) are homogeneous in the $x$ variables.

The definition and study of $G$ and its associated equation set are one of the central results of this work. The suggested approach is to associate with any Feynman diagram $\Gamma$ its numerator-free Lie Algebra, to study it and use it to appropriately reduce the corresponding Feynman integral as much as possible.

The name Integration By Parts is misleading from the perspective of $G$ since integration by parts is a general technique which applies to any integral whereas here the requirement to avoid numerators is essential and it reflect the diagram's topology.

We note that \cite{TarasovNum} showed that numerators can be eliminated by raising the dimension.


\subsection*{Constant-free sub-algebra}

Another useful sub-algebra is given by the following definition:

\presub {\bf Definition}. We denote by $G_2$ the subspace of $G$ such which is constant-free, namely such that $c^a=0$ in $D^a$ (\ref{def:set}).

$G_2$ is co-dimension 1 in $G$, it is a sub-algebra of $G$ and in fact an ideal, namely $[g,g_2] \in G_2$ for all $g \in  G,~ g_2 \in G_2$. For a vacuum diagram without numerators, the complement to $G_2$ is the equation \be
0= D_0\,  I \equiv \[ c_0 - x_i\, \del^i \] I 
\label{dim-eq}
\ee
where $c_0$ is the $x$ dimension of $I$. In other words, this equation is the Euler theorem for homogeneous functions reflecting dimensional analysis of $I$. We call $D_0:=c_0- x_i\, \del^i$ the \emph{dimension differential operator}.


\section{Outline for method of solution}
\label{sec:soln}


Given a vacuum Feynman diagram $\Gamma$ and a numerator-free integral (\ref{def:Ix}) we associated with it a set of equations (\ref{def:set}) and a numerator-free Lie algebra of differential operators (\ref{comm-relationsD}). In this section we outline the method of solution for this set. We reiterate that the current formulation is expected to generalize further to allow for external legs and integrals with numerators.

Any single equation or group generator, namely a specific index $a$ in the equation set, defines the standard field of characteristic curves $x_j=x_j(s)$ through \be
\frac{d}{ds} x_j = \(T^a\)^i_{~j}\, x_i
\label{curves}
\ee
These are first order equations, which in this case are simply linear with constant coefficients.

In the presence of a set of equations we have a set of characteristic curve fields.   This set generates characteristic sub-manifolds which foliate the $x$ space  (the co-dimension of the sub-manifolds could be greater than zero as a result of the closure of the commutation relations). In fact these sub-manifolds are nothing but orbits of the symmetry group $G$ in $x$ space.

It is useful to imagine coordinates which are adapted to the group action: $G$ orbits are parameterized by $\Xi^\alpha$, while $P_1,P_2,\dots$ are independent invariants of $G$ and can be used as coordinates for the transverse direction.


The equation set (\ref{def:set}) can be solved algebraically to yield the gradient of $I$ within the $G$-orbits as a function of $I$ and $x$: $\del/\del \Xi^\alpha I=f^\alpha(I,x)$. Therefore the computation of $I(x)$ can be reduced as follows \bi
\item Evaluate $I=I(P_1, P_2,\dots)$  on a freely chosen sub-manifold $X_0$ of initial conditions which is transversal to the $G$-orbits.
\item To evaluate $I(x)$ for any $x$, find $x_0 \in X_0$ on the same $G$-orbit as $x$, connect them by a curve $x^C=x^C(s)$, such that $x^C(0)=x_0, ~x^C(1)=x$ and finally solve an ordinary differential equation along the curve, namely \be
\frac{d}{ds} I = f^\alpha(I,x)\, \frac{dx^C_j}{ds} 
 \label{ode-reduct}
 \ee
\ei

Clearly \emph{the co-dimension of a $G$-orbit} is central to this approach as it sets the dimension of the initial value surface $X_0$. The freedom in choosing $X_0$ can be exploited to choose integrals with a higher degree of symmetry, for instance, several identical masses ($x$'s).

We can reduce the problem a bit further. Let us solve the homogeneous version of (\ref{def:set}), 
namely \be
0 = D^a\, I_0
 \label{def:homog-set}
\ee
If we consider only generators lying within the constant-free subgroup $G_2$ then by definition $I_0$ is constant on $G_2$ orbits. This leads us to split the coordinates $\Xi^\alpha$ into coordinates $\xi_\alpha$ which parameterize the $G_2$ orbit and one other coordinate, which we denote by $P$. Then we should consider the the dimension equation $D_0\, I_0=0$. It is an ODE in the dependent variable $P$, implying some solution $I_0=c(P_1,P_2,\dots)\, I_0(P)$.

Now solutions for the original, inhomogenous equations can be obtained as usual through the method of variation of the constants \be
I(x) = c(x)\, I_0 ~.
\ee
Upon substitution back into (\ref{def:set}) one finds the projected gradient of $c$, $\del/\del \xi_\alpha c = \tilde{f}^\alpha(x)$ where as usual $\tilde{f}^\alpha$ is now independent of $c$ and hence the solution for $c$ reduces to a line integral \be
c(x) = c(x_0) + \int_{\xi_0}^\xi \tilde{f}^\alpha(x)\, d\xi_\alpha
 \label{line-integ-reduct}
\ee

Finally I would like to mention another possible approach to solving system of equations (\ref{def:set}) over the $G$-orbits. One can consider not only first order differential equations, but also higher order differential equations. For example, one could consider a second order PDE formed from a quadratic Casimir of the differential operators $D^a$.

\section{Demonstration}
\label{sec:demonstrate}

In this section we shall analyze a specific Feynman integral with the method described above. But first we start with a more general perspective. In this paper we consider mostly vacuum diagrams and associated integral without numerators. It is illuminating to find the relation between the number of numerator terms and the number of loops.

\subsection*{Number of numerators}

A vacuum diagram with $L$ loops defines $Q=L(L+1)/2$ independent quadratic expressions (Lorentz scalars). The number of propagators $P$, on the other hand, assuming trivalent vertices is $P=3(L-1)$ [this is gotten from the following two equations: the Euler characteristic gives $L-P+V=1$ where $V$ is the number of vertices and the trivalent nature of the vertices implies $3 V=2 P$]. Hence the number of numerator terms is \be
\#Num = Q - P = \frac{L(L+1)}{2} - 3(L-1) = \half  (L-2) (L-3) 
\ee

Let us tabulate this function for some low values of $L$ \be \begin{tabular}{c|c|c|c|c|}
L 		&	2	&	3	&	4 &	5 \\
\hline 
\#Num	&	0	&	0	&	1 &  3
\end{tabular}
\label{number-num-table}
\ee

The IBP relations of a vacuum diagram are generated by the operators \be
O_{rs}=\del/\del l_r\, l_s
\label{def:O}
\ee
 which generate the algebra \be
 G_0=GL(L,\IR)
 \label{GLLR}
 \ee
 of invertible real $L$ by $L$ matrices. 

We shall now discuss some cases one by one.

\begin{figure}
\centering \noindent
\includegraphics[width=10cm]{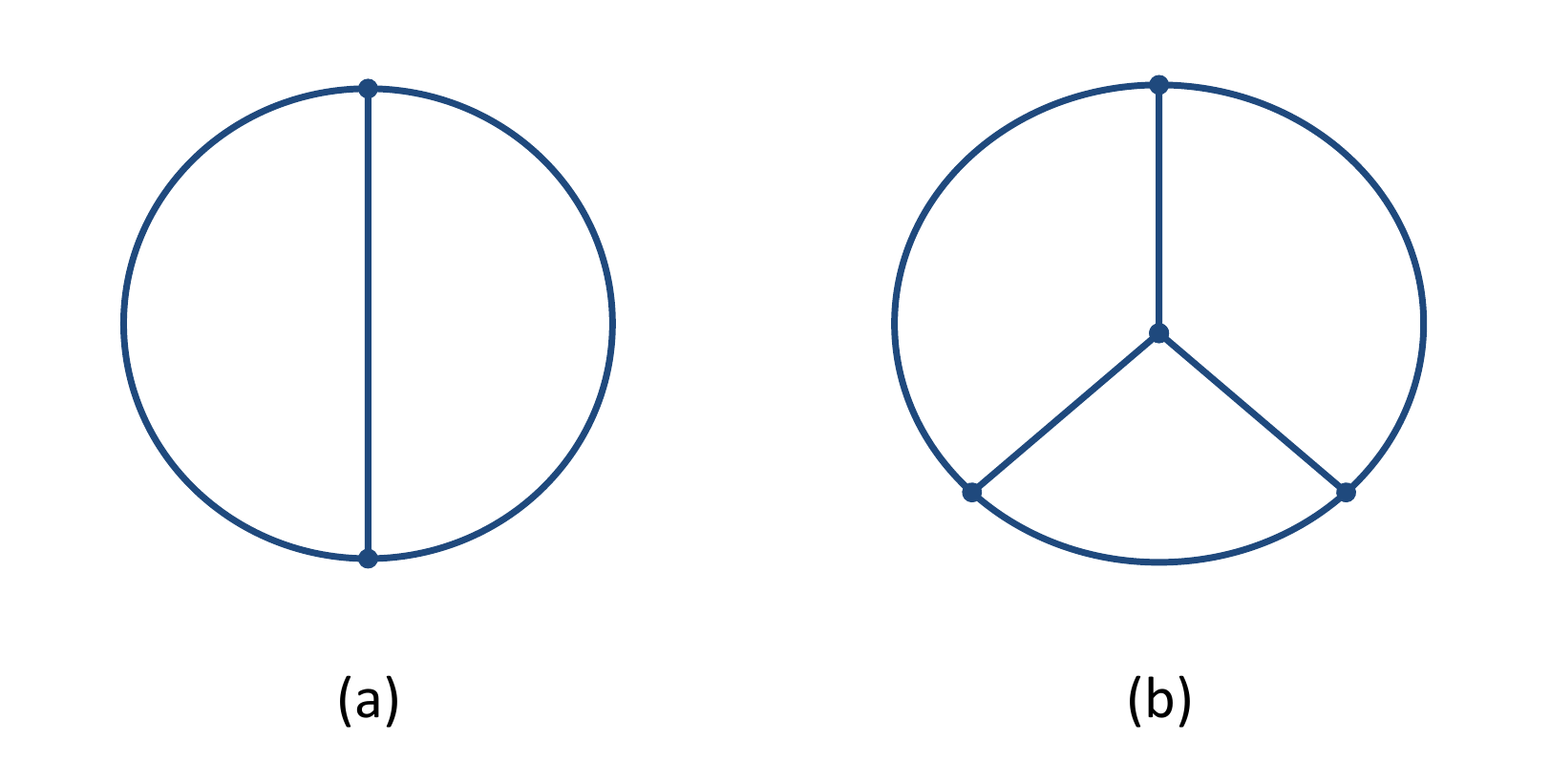}
\caption[]{Trivalent vacuum diagrams with few loops. (a) The 2-loop vacuum diagram. Its associated Feynman diagram is analyzed in the text. (b) The 3-loop trivalent vacuum diagram. It has a tetrahedron topology}
 \label{fig:vacuum}
\end{figure}

There is a single 2-loop trivalent vacuum diagram, see fig. \ref{fig:vacuum}(a). There are no numerators in this case, so all IBP relations are numerator-free and \be
G = G_0 = GL(2,\IR)  ~.
\ee

At 3-loops there is still a single such diagram -- the tetrahedron, see fig. \ref{fig:vacuum}(b). In this case again all IBP relations are numerator-free and so \be
G = G_0 = GL(3,\IR)  ~.
\ee

At 4-loops there are already 2 inequivalent trivalent vacuum diagrams, and there is a possible numerator. So we can expect that $G$ would be a proper subgroup of $GL(4,\IR)$.

Each non-vacuum diagram defines its ``vacuum closure"  in the usual way - by attaching all external legs to a new point at infinity.

\subsection*{Two-loop vacuum diagram}

We now turn to demonstrate the method by applying it to a specific diagram, the 2-loop (trivalent) vacuum diagram of fig. \ref{fig:vacuum}(a). It was computed in \cite{DavydychevTausk1992} using the Mellin-Barnes transform, see also references therein.

The symmetry group of the diagram allows to exchange the two vertices and to permute the 3 propagators and hence has 12 elements.

There are 3 propagators, so the generating function (\ref{def:Ix}) depends on 3 $x$ variables (mass-squares) $x_1,\, x_2,\, x_3$. According to table \ref{number-num-table} there are no possible numerators, and so this function generates all the possible Feynman integrals. For a specific choice of loop currents its definition becomes \be
I\(x_1,\, x_2,\, x_3\) = \int \frac{d^dl_1\, d^dl_2}{\( l_1^2-x_1\) \( l_2^2-x_2\) \( (l_1+l_2)^2-x_3\)}
\label{def:Ix-Ln20}
\ee

The set of all possible IBP's generates the group $G_0=GL(2,\IR)$ according to (\ref{GLLR}) with $L=2$. As there are no possible numerators, there are no possible obstructions for the definition of the numerator-free sub-algebra, and we have $G=G_0= GL(2,\IR)$.

\begin{figure}[t!]
\centering \noindent
\includegraphics[width=5cm]{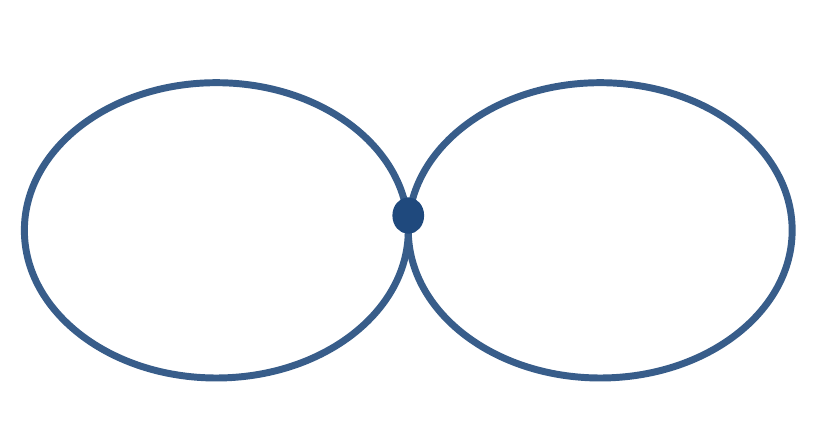}
\caption[]{The one and only possible degeneration of fig. \ref{fig:vacuum}(a).}
\label{fig:fig8}
\end{figure}

In order to form the equation set (\ref{def:set}) we need to know the Feynman integrals of all possible degenerations. In this case there is a single degeneration into the figure 8 diagram shown in fig. \ref{fig:fig8}, whose value factorizes \be
I _{\infty}(x_1,x_2) = j(x_1)\, j(x_2) 
\label{def:j}
\ee
and can be considered to define the function $j$. 

The equation set can now be cast in the following rather symmetric form  \be
 0 = \[ \begin{array}{cccc}
 d-3 	& -x_1		& -x_2		& -x_3 	\\
 0 		& x_2-x_3	& x_2		& -x_3 	\\
 0 		& -x_1 		& x_3-x_1	& x_3	\\
 0		& x_1 		-x_2		& x_1- x_2 \\
 \end{array} \] \,
 \[ \begin{array}{c}
1 \\
\del_1 \\
\del_2 \\
\del_3 \\
\end{array} \] \, I +
\[ \begin{array}{c}
0 \\
(j_3-j_2)\, j_1' \\
(j_1-j_3)\, j_2' \\
(j_2-j_1)\, j_3' \\
\end{array} \]
\label{def:set-Ln20}
\ee
where the functions $j_i$ which appear in the sources are given by \be
j_i := j(x_i) 
\label{def:ji}
\ee

The first row can be identified with the dimension generator $D_0$ (\ref{dim-eq}), while the next three rows generate the group $G_2=SL(2,\IR) \subset G$. More explicitly, the first generator is \be
L_1 = (x_2-x_3) \del^1+ x_2\, \del^2 - x_3\, \del^3
\ee
and $L_2,L_3$ are defined by cyclic permutations.  The commutation relations among the $L_i$'s  are \be
\[ L_1,L_2 \] = -L_1- L_2
\ee
and the rest can be obtained by cyclic permutations.

We confirmed that the sources satisfy the integrability constraint (\ref{integ-constraint}).

The $x$ parameters transform in the same way as the quadratics $l_r \cdot l_s$, so they are in the symmetric representation, which indeed has dimension 3. In fact we can define a symmetric matrix in terms of $x$ \be
x_{rs}=\[ \begin{array}{cc}
2 x_1 	& -x_1-x_2 + x_3 \\
-x_1-x_2 + x_3 & 2 x_2 \\
\end{array} \]
\ee 
then $g \in G=GL(2,\IR)$ acts on it through \be
x_{rs} \to g\, x_{rs}\, g^T 
\label{x-transformation}
\ee 

The orbits of $G$ are 3-dimensional so the initial conditions sub-manifold is 0-dimensional, namely a point. In fact, the integral is known at the following two points $x_1=x_2=0,\, x_3=m^2$ (e.g. fig 3.4 on p.37 and eq. (10.39) of \cite{SmirnovBook2012}) as well as $x_1=0, \, x_2=x_3=m^2$ (e.g. example 3.4 on p. 48 and eq. (10.38) of \cite{SmirnovBook2012}).

In order to obtain homogenous solutions for (\ref{def:set-Ln20}) we notice that there is a single $G_2$ invariant $x$ expression, namely \be
P := {\rm det} \( x_{rs} \) = \(x_1 + x_2 + x_3 \)^2  -2 \(x_1^2 + x_2^2 + x_3^2\) 
\label{def:P}
\ee
We notice that $P$ is symmetric in $x_i$, as it should, even though the matrix $x_{rs}$ was not so, due to the breaking of the symmetry by the choice of loop currents. This is the same polynomial that appears in Baikov's method \cite{Baikov1996} when applied to this case, see \cite{SmirnovBook2006} eq. (6.17) after the replacement $q^2 \to x_3$.

The first row of (\ref{def:set-Ln20}) allows now to solve for $I_0$ to obtain \be
I_0 = const\, P^{(d-3)/2}
\ee

Defining \be
I(x) = c(x) I_0(P)
\ee
we can solve (\ref{def:set-Ln20}) to obtain the gradient of $c$ and thereby reduce the solution of $I(x)$ to a line integral.

Alternatively we may consider the 2nd order differential equation given by the quadratic Casimir \be
\Delta = \( L_1 + L_2 + L_3 \)^2- \( L_1^2 +  L_2^2 + L_3^2 \) 
\ee



\section{Summary and Discussion}
\label{sec:summary}






\noindent {\bf Summary}. Our goal is to fully exploit the IBP method to compute Feynman integrals. The standard method is to employ recursion relations to reduce the given integral to a set of master integrals using a computerized algorithm and then to compute the master integrals, possibly through the method of differential equations.

We find that the generating function $I=I(x)$ defined in (\ref{def:Ix}) such that $x$ denotes mass-squared, satisfies a set of differential equations (4.7) in $x$ space  indexed by the generators of a Lie group $G$. The group $G$ appears to be a new concept. It  depends only the diagram's topology $G \equiv G_\Gamma$, where $\Gamma$ represents the diagram, and can be used to characterize it.  The group $G$ foliates $x$ space into $G$ orbits. The PDE set reduces the computation of the general Feynman integral to the evaluation of the Feynman integral over an initial value sub-manifold of parameters which is transversal to the $G$-orbits, followed by the evaluation of a line integral. Another approach using higher order differential equations was suggested.

The method was demonstrated for the case of the 2-loop vacuum diagram. This diagram consists of  3 propagators and hence we have 3 $x$'s: $x_1,\, x_2,\, x_3$ . 
The initial group $G_0$ is $G_0=GL(L,\IR)=GL(2,\IR)$. All IBP relations are numerator-free so the group $G=G_0=GL(2,\IR)$ and the constant-free group is $G_2=SL(2,\IR)$. The orbits of $G$ are 3d (co-dimension 0). Hence the surface of initial conditions is a single point $x_0$ to be chosen at will. In fact there two are points $x_0$ where $I(x_0)$ is known. The orbits of $G_2$ are given by the level-sets of the Baikov polynomial $P=(x_1+x_2+x_3)^2 -2 (x_1^2 + x_2^2 + x_3 ^2)$. The homogeneous solutions are $I_0 = const\, P^{(d-3)/2}$. In this way the evaluation of $I(x)$ for general $x$ is reduced to the evaluation at a single point $I(x_0)$ and  a line integral.

\subsection{Discussion}

\noindent {\bf Interpretation of $G$}. The IBP relations define a Lie algebra, which is particularly evident in $x$ space. This algebra generates a group which defines the $G$-orbits in $X$ space. The $G$-orbits describe the reach of the set of differential equations - they enable to replace the required point $x$ by any point $x_0$ lying on the same $G$-orbit. 

\presub {\bf The `instruction manual' and the $Y-\Delta$ transform}. So far we did not address the original question, namely when the method should be applied and what the expected  benefits should be. We now see that the method is more useful the larger $G$ is, and more precisely the smaller is the co-dimension of the $G$-orbits.  There remains the general question how $G$ is to be determined given $\Gamma$.

I would like to speculate further on the diagrams where the method applies. The classic IBP relation is the triangle identity. This can be related to a known concept from the field of electric circuits. The Kirchhoff equations are known to determine a solution for any circuit. However, in certain cases the solution can be obtained through a stepwise reduction, namely when components are connected either in parallel or in series. In fact, there is another case where stepwise reduction is possible, namely the case of the triangle which can be transformed into a Y shaped sub-circuit through the so-called $Y-\Delta$ transform. We would like to conjecture that a Feynman integral (without numerators) can be fully solved in $n$ space and expressed in terms of $\Gamma$ functions alone exactly when the corresponding circuit is reducible using the $Y-\Delta$ transform. For example, out of the three 3-loop propagator diagrams L, M and N only N, the non-planar one cannot be reduced by IBP to $\Gamma$ functions, and only its associated electric circuit cannot be stepwise reduced not even through the use of the $Y-\Delta$ transform.

\presub {\bf Relation of IBP and DE methods}. The methods of IBP and DE and known to be related. From the current perspective, they in fact appear to be one, as the IBP relations are represented by differential equations in the space of $x$ parameters.

\presub {\bf Relations with the literature}. In our demonstrated example the homogenous solution $I_0(x)$ turns out to be the same as Baikov's function $g(x)$. This is likely not a coincidence. 

In the usual method it is important and interesting to know the number of master integrals. In 2010 this number was shown to be finite \cite{SmirnovPetukhov2010}. In 2013 it was shown in \cite{LeePomeransky2013} to be counted by certain critical points of  the Kirchhoff-Symanzik polynomials. Clearly this number depends only on the diagram's topology and one can hope for an even simpler method to obtain such a simple integer. This question in fact was a main original motivation to the current research. The current results take away emphasis from master integrals and their number, and instead suggest to focus on other discrete data, including the identity of $G$ and the co-dimension of its $G$-orbits.

\presub {\bf Open questions}. I see many open questions, several of them are currently under study.
\subsection*{Acknowledgments}

I would like to thank V. Smirnov for very helpful correspondence and Ruth Shir for collaboration on a related project.

This research was supported by the Israel Science Foundation grant no. 812/11 and it is part of the Einstein Research Project "Gravitation and High Energy Physics", which is funded by the Einstein Foundation Berlin.

\bibliographystyle{unsrt}

\end{document}